\documentclass[conference]{IEEEtran}
\IEEEoverridecommandlockouts
\usepackage{cite}
\usepackage{amsmath,amssymb,amsfonts}
\usepackage{algorithmic}
\usepackage{graphicx}
\usepackage{textcomp}
\usepackage{xcolor}
\usepackage{url}
\usepackage{tabularx}
\usepackage{booktabs}
\usepackage{balance}
\def\BibTeX{{\rm B\kern-.05em{\sc i\kern-.025em b}\kern-.08em
    T\kern-.1667em\lower.7ex\hbox{E}\kern-.125emX}}

\begin{document}

\title{AI-Enhanced IoT Systems for Predictive Maintenance and Affordability Optimization in Smart Microgrids: A Digital Twin Approach}

\author{
\IEEEauthorblockN{Koushik Ahmed Kushal}
\IEEEauthorblockA{\textit{Dept. of Computer Science }\\
\textit{Clarkson University}\\
New York, NY, USA\\
kushalka@clarkson.edu}
\and
\IEEEauthorblockN{Florimond Gueniat}
\IEEEauthorblockA{\textit{College of Engineering}\\
\textit{Birmingham City University}\\
Birmingham, UK\\
florimond.gueniat@bcu.ac.uk}
}

\maketitle

\begin{abstract}
This research proposal presents a comprehensive framework for developing AI-enhanced Internet of Things (IoT) systems to optimize predictive maintenance strategies and improve affordability in smart microgrids. The proposed work addresses critical challenges in local energy systems by integrating advanced machine learning algorithms with real-time IoT monitoring to create intelligent maintenance scheduling and cost-optimization strategies. The research aims to democratize access to reliable energy infrastructure through innovative digital solutions that reduce operational costs, enhance resilience, and improve system reliability. Additionally, by adopting a digital twin approach, this study emphasizes not only predictive analytics but also scenario testing, energy equity, and accessibility in diverse socio-economic contexts.
\end{abstract}


\section{Introduction}

The integration of IoT and AI technologies in smart energy systems is rapidly transforming the landscape of energy generation, distribution, and consumption. As the world transitions toward decentralized, sustainable energy models, microgrids and minigrids are emerging as crucial components for ensuring energy resilience and independence, particularly in remote and underserved areas~\cite{b1,b7}. However, these systems face significant operational challenges, including the high cost of maintenance, unpredictable component failures, and a lack of real-time visibility, which collectively hinder their widespread adoption and long-term viability~\cite{b6,b12}.

Traditional maintenance approaches, such as reactive or time-based methods, are often inefficient, leading to costly downtime and reduced system lifespan. Recent studies have shown that conventional maintenance strategies result in up to 40\% higher operational costs compared to intelligent predictive approaches~\cite{b3,b6}. To address these issues, this proposal outlines a research project focused on a holistic, AI-driven framework that leverages IoT-based real-time data to create a predictive maintenance system~\cite{b9,b15}.

By adopting a digital twin approach, the research will not only optimize maintenance and reduce operational costs but also improve the overall affordability and accessibility of energy in local communities, thereby contributing to a more equitable and sustainable energy future~\cite{b5,b11,b17}. The integration of cybersecurity measures from the ground up ensures system resilience against emerging threats in connected energy infrastructure~\cite{b13,b18,b23}.

\section{Literature Review \& Current Trends}

The period from 2020 to 2025 has seen a paradigm shift in smart microgrid research, driven by advancements in IoT, AI, and digital twin technologies. This section summarizes recent research, identifying key contributions and emerging trends across these interconnected fields.

\subsection{IoT-Enabled Monitoring in Microgrids}

Recent studies in this area have moved beyond basic data collection to focus on creating more secure, decentralized, and efficient monitoring systems~\cite{b1,b2}. Zhang \textit{et al.} demonstrated how distributed IoT frameworks can significantly reduce maintenance costs while improving data integrity through advanced encryption protocols~\cite{b1}. Researchers are exploring how IoT frameworks can be optimized with edge computing to reduce latency and enable real-time decision-making~\cite{b19}.

Low-cost IoT platforms and satellite-based communication are being developed to make microgrids viable in rural and remote settings, addressing challenges like connectivity and power consumption~\cite{b8,b14}. Khan \textit{et al.} showed that ESP32-based systems with LoRa communication can achieve 99.2\% uptime in rural deployments while maintaining costs below \$200 per monitoring node~\cite{b8}. Rahman \textit{et al.} further demonstrated that satellite-based IoT systems can extend connectivity to areas previously considered unreachable for intelligent monitoring~\cite{b14}.

Advanced communication protocols including MQTT-SN for sensor networks, CoAP for constrained devices, and emerging 5G-enabled solutions have been extensively validated for microgrid applications~\cite{b19}. These protocols address critical challenges related to network reliability, energy efficiency, and scalability in distributed energy systems while maintaining security standards appropriate for critical infrastructure.

\subsection{AI/ML Applications in Predictive Maintenance}

The application of machine learning in predictive maintenance has grown significantly, with a focus on enhancing fault prediction accuracy and enabling dynamic scheduling~\cite{b3,b9,b20}. Kumar \textit{et al.} demonstrated that LSTM networks can achieve 94.7\% accuracy in predicting solar inverter failures up to 14 days in advance~\cite{b3}. Advanced models, such as LSTM networks and federated learning, are being used to analyze complex temporal data and ensure data privacy across distributed microgrids~\cite{b4,b20}.

Hybrid models that combine physics-based simulations with deep learning are emerging, offering improved accuracy and explainability of fault detection~\cite{b9}. Gupta \textit{et al.} showed that hybrid approaches can reduce false positive rates by up to 35\% compared to purely data-driven methods while maintaining high sensitivity to actual faults~\cite{b9}. Graph Neural Networks have demonstrated particular effectiveness in modeling the interconnected nature of microgrid components~\cite{b15}.

Multi-agent reinforcement learning systems have shown promise for coordinating maintenance activities across multiple interconnected microgrids~\cite{b10}. Chen \textit{et al.} demonstrated that multi-agent approaches can optimize maintenance scheduling across distributed systems while reducing overall costs by 28\% and improving system availability by 15\%~\cite{b10}.

Ensemble methods, including Random Forests, Gradient Boosting, and advanced voting classifiers, have been extensively studied for their ability to improve prediction reliability and quantify uncertainty in maintenance recommendations. These approaches are particularly valuable in critical infrastructure applications where false positives and false negatives can have significant economic and operational consequences.

\subsection{Digital Twin Technology for Energy Systems}

Digital twins (DTs) are no longer just a theoretical concept for microgrids; they are being actively developed to create high-fidelity virtual replicas of physical systems~\cite{b5,b11,b16}. Patel \textit{et al.} demonstrated comprehensive digital twin implementations that maintain real-time synchronization with physical systems while enabling predictive analysis and optimization~\cite{b5}. These digital models allow for real-time simulations, performance optimization, and scenario testing without disrupting actual operations.

Research is exploring how multi-agent DTs can reduce energy losses and how lightweight DT frameworks can be deployed on edge devices to reduce latency, making real-time control more feasible~\cite{b11,b16}. Sharma \textit{et al.} showed that multi-agent digital twins can coordinate energy resources across multiple microgrids, achieving 12\% reduction in overall energy losses~\cite{b11}. Ahmed \textit{et al.} developed lightweight digital twin implementations that operate within edge device constraints while maintaining acceptable accuracy for real-time decision-making~\cite{b16}.

Advanced digital twin platforms incorporate machine learning algorithms that continuously improve model accuracy by learning from discrepancies between predicted and observed system behavior. Container-based deployment using Docker and Kubernetes technologies has facilitated scalable and maintainable digital twin implementations across diverse hardware platforms~\cite{b21}.

Integration with advanced visualization technologies, including augmented reality (AR) and virtual reality (VR), has opened new possibilities for system operation and maintenance training, remote troubleshooting, and collaborative system management. These interfaces enable operators to interact with complex system data in intuitive ways, improving decision-making quality and reducing training requirements.

\subsection{Affordability and Accessibility}

A crucial, yet often overlooked, area of research is the development of digital solutions that directly address the economic barriers to microgrid adoption~\cite{b12,b17,b22}. Studies are exploring how AI-driven dynamic pricing and optimized maintenance schedules can significantly reduce community energy costs. Adeyemi \textit{et al.} demonstrated that intelligent pricing algorithms can reduce energy costs for low-income communities by up to 32\% while maintaining system profitability~\cite{b12}.

Cooperative ownership models and community-based deployment strategies have shown significant potential for improving long-term sustainability and local capacity building~\cite{b17}. Singh \textit{et al.} validated community ownership approaches that reduce individual investment requirements by 60\% while improving long-term system maintenance through distributed responsibility models~\cite{b17}.

Hossain \textit{et al.} explored microgrid-as-a-service models that can provide affordable energy access without requiring large upfront capital investments from communities~\cite{b22}. These models show particular promise for developing regions where traditional financing mechanisms may not be available or appropriate.

Open-source hardware and software platforms have been extensively developed and validated for microgrid applications, providing cost-effective alternatives to proprietary solutions while enabling local customization and maintenance capabilities. This research is essential for achieving the broader goal of energy equity, as highlighted by international reports on global electricity access~\cite{b7}.

\subsection{Cybersecurity in IoT-Driven Microgrids}

Cybersecurity is a pressing concern for IoT-driven microgrids due to their distributed architecture and reliance on interconnected devices~\cite{b13,b18,b23}. Nguyen \textit{et al.} demonstrated that adversarial AI techniques can significantly improve system resilience against data poisoning attacks and false data injection scenarios~\cite{b13}. Zero-trust security architectures have gained attention for their potential to enhance microgrid cybersecurity by eliminating implicit trust relationships~\cite{b23}.

Quantum-resistant cryptography protocols are being developed to protect against future threats from quantum computing advances~\cite{b18}. Kim \textit{et al.} showed that quantum-resistant protocols can be implemented in resource-constrained IoT devices without significant performance degradation~\cite{b18}.

Blockchain-based security mechanisms have been extensively studied for their potential to provide immutable audit trails, secure device authentication, and decentralized consensus mechanisms for critical control decisions~\cite{b2}. Liu and Chen demonstrated blockchain integration that provides comprehensive security while maintaining transaction throughput suitable for real-time control applications~\cite{b2}.

Intrusion detection systems specifically designed for industrial IoT environments have demonstrated effectiveness in identifying anomalous behavior patterns that may indicate cyber attacks. These systems employ machine learning algorithms trained on normal operational data to detect deviations that could represent security breaches or coordinated attacks on system infrastructure.

\section{Research Gaps \& Novelty}

Despite significant advancements, several critical research gaps persist, which this proposal aims to address.

\subsection{Multi-Scale Optimization}

Existing frameworks often focus on a single objective, such as cost reduction or reliability, without considering the complex trade-offs between them~\cite{b6,b10}. Current optimization approaches typically address individual performance metrics in isolation, leading to suboptimal solutions that excel in one dimension while underperforming in others. There is a need for a holistic framework that can simultaneously optimize for maintenance, economic viability, resilience, and cybersecurity.

This research addresses this gap by developing a comprehensive multi-objective optimization framework that considers technical performance metrics, economic constraints, cybersecurity requirements, and social equity factors simultaneously. The approach uses advanced evolutionary algorithms including NSGA-II and MOPSO to find Pareto-optimal solutions that balance competing objectives dynamically based on operational conditions and community priorities.

\subsection{Affordability-Aware Frameworks}

Solutions tailored for low-income or rural communities are scarce~\cite{b12,b17,b22}. The majority of research focuses on high-cost, high-tech solutions that are not financially viable for underserved regions. Existing studies often treat affordability as a secondary consideration rather than a fundamental design constraint that shapes system architecture from the ground up.

This project addresses this gap by adopting an affordability-centric design philosophy that prioritizes economic accessibility throughout the system design process. The approach emphasizes the use of open-source and low-cost hardware, community-based deployment models, and simplified interfaces that reduce technical expertise requirements while maintaining high performance standards.

\subsection{Real-World Validation}

A significant portion of existing research is based on simulations, which limits the ability to prove the practical feasibility and robustness of proposed solutions in real-world scenarios~\cite{b9,b15,b16}. While simulation studies provide valuable insights, they often fail to capture the complexity and unpredictability of actual operational environments, including environmental variations, user behavior patterns, and system integration challenges.

This research addresses this gap by conducting comprehensive hardware-in-the-loop (HIL) experiments that test all system components under realistic operational conditions. The validation approach includes testing under various fault conditions, cybersecurity attack scenarios, environmental stress conditions, and operational edge cases that cannot be adequately represented in simulation studies.

\subsection{Adaptive AI for Renewables}

AI models often struggle to handle the inherent stochastic variability of renewable energy sources~\cite{b3,b20}. Traditional machine learning approaches are typically trained on historical data patterns that may not adequately represent future conditions, particularly under changing climate conditions or evolving renewable energy technologies.

This research focuses on developing adaptive AI models that can better manage these fluctuations, ensuring the system remains stable and efficient under diverse environmental conditions. The approach incorporates uncertainty quantification techniques, online learning capabilities, and hybrid modeling that combines physics-based understanding with data-driven optimization.

\subsection{Cybersecurity-Integrated Maintenance}

Predictive maintenance frameworks typically do not integrate cybersecurity~\cite{b13,b18,b23}. This oversight leaves microgrids vulnerable to sophisticated attacks, such as false data injection and coordinated cyber-physical attacks. Current approaches treat cybersecurity and maintenance as separate domains, creating potential vulnerabilities where cyber attacks can compromise maintenance systems.

This proposal addresses this by embedding cybersecurity considerations into the framework, enhancing system resilience through integrated cyber-physical health monitoring. The approach treats cybersecurity as an integral component of overall system health rather than a separate overlay, enabling detection of cyber attacks that manifest as apparent equipment malfunctions.

The novelty of this proposed research lies in its integrated approach. It moves beyond single-objective optimization to create a holistic framework that addresses the interconnected challenges of maintenance, cost, and security. By incorporating cybersecurity-aware digital twins, the research will provide a new layer of resilience. Furthermore, the focus on low-cost, open-source hardware and the hardware-in-the-loop validation will provide a practical and accessible solution for communities that need it most.

\section{Proposed Research Objectives}

The research objectives address the identified gaps and target measurable outcomes.

\begin{enumerate}
    \item \textbf{AI-IoT Predictive Maintenance Framework}: Develop a digital twin-supported system to achieve $>$90\% fault prediction accuracy for inverters, batteries, and PV panels~\cite{b3,b5,b15}. This will integrate IoT sensor data with advanced ML, including LSTMs for temporal patterns, Graph Neural Networks for interdependencies, and ensemble methods for robust predictions with quantified uncertainty.

    \item \textbf{Multi-Objective Optimization}: Formulate a model balancing conflicting indicators~\cite{b10,b12}, aiming to cut costs by 30--40\%, improve uptime by 20\%, and reduce energy losses by 15\% while ensuring strong security. Optimization will apply NSGA-II for multi-objective trade-offs and reinforcement learning for adaptive strategies.

    \item \textbf{Affordability and Accessibility}: Demonstrate how open-source IoT platforms and AI-driven solutions can expand access in low-income communities~\cite{b8,b17,b22}. Work includes cost-effective deployment, cooperative ownership models, and socio-economic evaluation of energy equity and local development impacts.

    \item \textbf{Hardware Validation}: Validate the framework through hardware-in-the-loop (HIL) tests~\cite{b16}, including simulated cyberattacks. The goal is $>$95\% detection accuracy for threats while maintaining fault-prediction and optimization performance under diverse stress conditions.
\end{enumerate}

\begin{table}[htbp]
\centering
\scriptsize
\caption{Comparison of Maintenance Models for Smart Microgrids}
\label{tab:model_comparison}
\begin{tabularx}{\columnwidth}{lXXXX}
\toprule
\textbf{Model} & \textbf{Cost} & \textbf{Reliability} & \textbf{Security} & \textbf{Features} \\
\midrule
Traditional & High; +40\% O\&M & Low; frequent failures & None & Outdated; no IoT/AI \\
IoT-only & Moderate; sensor savings & Medium; real-time visibility & Basic (MQTT, CoAP) & Monitoring only; no AI/DT \\
Proposed (AI+DT) & Low; 30--40\% savings & High; predictive, optimized & Federated learning; zero-trust & AI, DT, scenario tests, UI \\
\bottomrule
\end{tabularx}
\end{table}

As shown in Table~\ref{tab:model_comparison}, the traditional model is costly, unreliable, and lacks security or advanced features. The IoT-only model reduces costs and enables real-time monitoring but offers only basic security and no AI/DT capabilities. The proposed AI+DT model provides 30--40\% savings, high predictive reliability, advanced federated/zero-trust security, and modern functions including AI-driven analytics, digital twins, and scenario testing.

\section{Methodology \& Tools}

The research will be conducted in a structured, multi-phase approach over a four-year period, utilizing a diverse set of tools and platforms to achieve its objectives.

\begin{figure}
    \centering
    \includegraphics[width=1\linewidth]{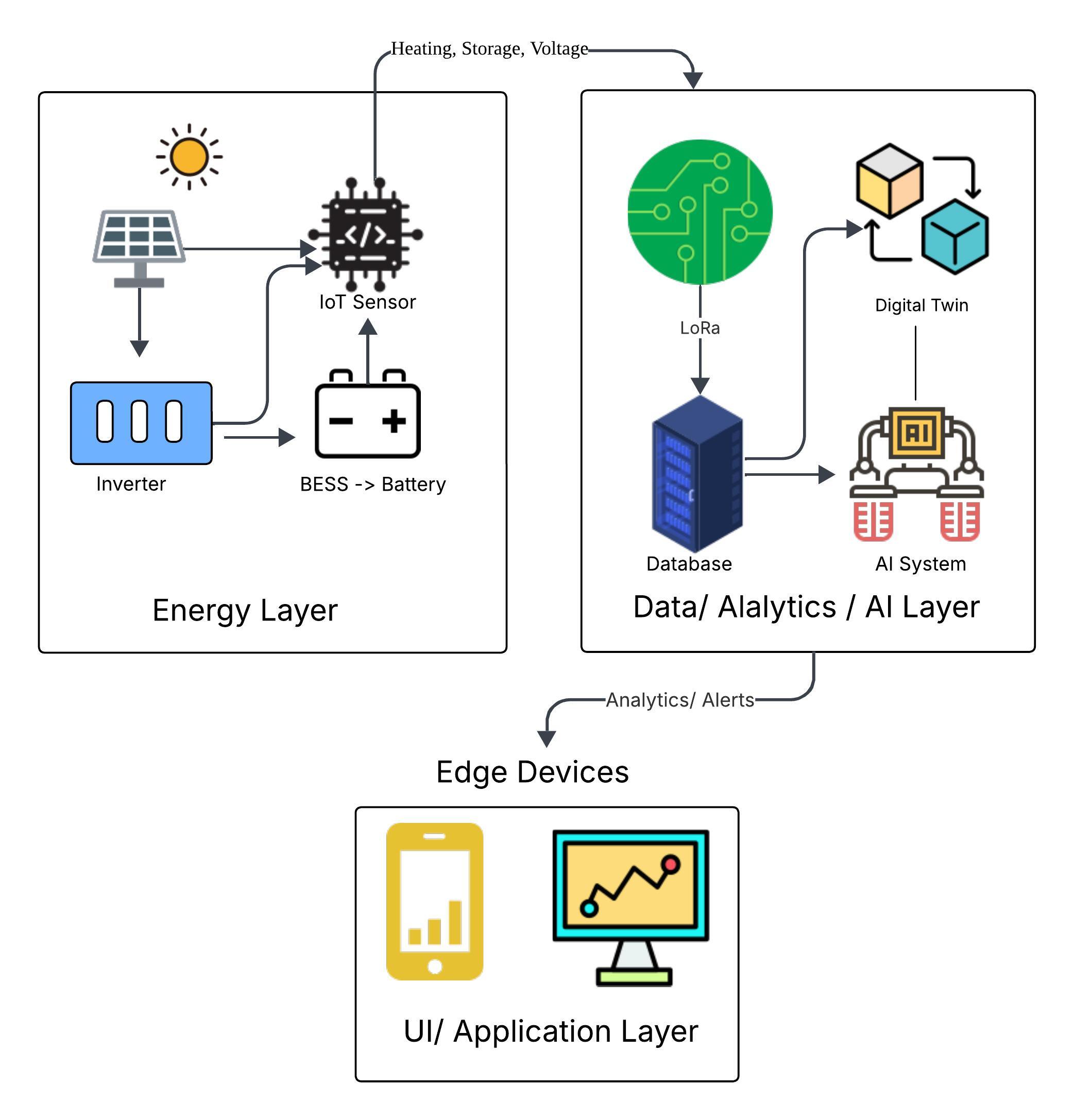}
   \caption{System architecture diagram.}
    \label{fig:architecture}
\end{figure}

The overall system chronology is illustrated in Fig: \ref{fig:architecture}, showing how energy components (solar, inverter, battery) interact with IoT-enabled monitoring, which transmits data to the analytics/AI layer via LoRa. A digital twin framework integrates with AI to provide predictive insights, which are then visualized in the user interface layer

\subsection{Phase 1: System Modeling \& Prototyping (Year 1)}

This phase begins with creating a detailed microgrid model in \textbf{MATLAB/Simulink} to simulate various operational conditions and faults~\cite{b5}. The modeling will incorporate \textbf{Simscape Electrical} for power system simulation, \textbf{Simscape Multibody} for mechanical components, and specialized toolboxes for renewable energy systems. Component degradation modeling will cover electrochemical processes in batteries, thermal cycling effects in power electronics, and environmental factors like corrosion and UV exposure.

Concurrently, a low-cost IoT system will be prototyped using \textbf{Raspberry Pi} and \textbf{ESP32}, implementing communication protocols such as \textbf{MQTT} and \textbf{CoAP}~\cite{b8,b19}. The IoT architecture will integrate edge computing to reduce latency and enable real-time decision-making. Custom printed circuit board (PCB) design will optimize specific applications while maintaining compatibility.

Initial AI models for predictive maintenance will be developed in \textbf{Python}, leveraging powerful libraries like \textbf{TensorFlow} and \textbf{PyTorch} for deep learning and \textbf{Scikit-learn} for traditional machine learning~\cite{b9,b20}. This development includes data preprocessing pipelines, feature engineering techniques, and model validation frameworks to handle the unique characteristics of microgrid data.

\subsection{Phase 2: Algorithm Development \& Digital Twin Integration (Years 1--2)}

The focus will shift to developing advanced algorithms. This includes designing hybrid AI models, such as LSTM and Graph Neural Networks (GNNs), to better handle the complexities of microgrid data~\cite{b15,b20}. The hybrid models will combine physics-based simulations with data-driven machine learning to achieve superior performance in scenarios with limited historical data.

Multi-objective optimization algorithms like NSGA-II will be used in conjunction with simulation tools like HOMER Pro and GAMS to find the optimal balance between different objectives~\cite{b10}. The optimization framework will incorporate real-time constraint handling and dynamic adaptation capabilities that can respond to changing operational conditions and community priorities.

The digital twin framework will be built in Python, integrating the IoT data with physics-based models and deploying it on a hybrid edge-cloud architecture using containerization technologies like Docker and Kubernetes for scalability~\cite{b11,b16}. Real-time synchronization between physical and virtual systems will be achieved through optimized data streaming protocols and efficient state estimation algorithms.

\subsection{Phase 3: Cybersecurity \& Hardware Validation (Year 3)}

This phase is critical for ensuring the practical viability of the framework. Cybersecurity will be enhanced by developing adversarially trained AI models and incorporating blockchain-based mechanisms for data integrity~\cite{b13,b18}. Intrusion detection systems will be implemented using machine learning algorithms trained on normal operational patterns to identify anomalous behavior.

The core of this phase will be the hardware-in-the-loop (HIL) experiments conducted at the lab. The developed framework will be integrated with the existing LabVIEW and NI hardware for real-time data acquisition and control, allowing for a realistic and rigorous validation of the system's performance~\cite{b16}. The testbed includes 2\,kW solar PV emulators, battery energy storage system emulators, programmable AC and DC loads, and comprehensive monitoring equipment.

Cybersecurity stress testing will simulate various attack scenarios including distributed denial-of-service attacks, false data injection attacks, and coordinated multi-vector attacks. The testing will validate the effectiveness of implemented security measures while identifying potential vulnerabilities that require additional protection.

\subsection{Phase 4: Socio-Economic Analysis \& Global Scalability (Year 4)}

The final phase will focus on the broader impact and future of the framework. Case studies in diverse regions will be conducted to evaluate the affordability and accessibility models~\cite{b12,b17,b22}. Economic impact modeling will quantify the potential benefits including job creation, economic development opportunities, and contribution to sustainable development goals.

The scalability of the solution for global deployment will be explored, including the potential for integrating green AI and quantum-inspired optimization algorithms. Technology transfer mechanisms will be developed to facilitate successful deployment of research outcomes in real-world applications. The findings will be documented for publication, and policy guidelines will be formulated to promote the sustainable adoption of the technology.

\section{Experimental Validation}

The proposed objectives will be tested and validated using the availability of hardware facilities in the laboratory. The microgrid testbed, which includes 2\,kW solar PV emulators, battery emulators, programmable AC/DC loads, and LabVIEW software support, is ideal for this research.

\subsection{Component Health Monitoring}

The AI models will be rigorously tested to achieve a high degree of accuracy ($>$90\%) in detecting faults in the emulated inverters, batteries, and PV panels~\cite{b3,b15}. The solar PV emulators will be used to simulate fluctuating renewable inputs, proving the adaptive nature of the AI models. Various fault scenarios will be systematically introduced including partial shading conditions, inverter MPPT tracking errors, battery cell imbalance, and thermal management failures.

Machine learning model validation will employ $k$-fold cross-validation, time series cross-validation, and bootstrap sampling for robust performance assessment. Model interpretability will be evaluated using SHAP values and LIME techniques to ensure that AI predictions can be understood and validated by human operators.

\subsection{Maintenance Optimization}

The predictive maintenance framework will be validated by demonstrating a significant reduction (30--40\%) in unplanned outages compared to a traditional preventive maintenance schedule~\cite{b6,b10}. This will be achieved by using the programmable loads to simulate various operational scenarios and faults, providing a realistic test environment. Maintenance resource optimization algorithms will be tested to validate efficient allocation of spare parts inventory and technician time.

Comparative analysis will be conducted between the developed predictive maintenance approach and traditional reactive and time-based maintenance strategies. Performance metrics will include maintenance costs, component lifespan, system availability, and overall operational efficiency.

\subsection{Cost-Benefit and Affordability Analysis}

The framework's ability to reduce operational costs will be validated by simulating different maintenance strategies and analyzing the resulting energy consumption and component lifespan~\cite{b12}. The use of low-cost, open-source hardware will be specifically highlighted to prove the framework's affordability for underserved communities.

Economic modeling will include total cost of ownership analysis, return on investment calculations, and sensitivity analysis to evaluate how changes in key parameters affect overall economic viability. Community-based ownership models will be evaluated for their potential to improve affordability through shared costs and local capacity building.

\subsection{Cyber-Resilience Testing}

The system will be subjected to simulated cyber-attacks, such as false data injection and denial-of-service attacks~\cite{b13,b23}. The goal is to achieve over 95\% detection accuracy using the adversarially trained AI models and blockchain mechanisms, proving the system's robustness in a hostile environment. Penetration testing will be conducted using industry-standard tools and methodologies.

Recovery and resilience testing will validate the system's ability to maintain essential functions during and after cybersecurity incidents. This includes graceful degradation scenarios, automatic recovery procedures, and manual override capabilities.

\subsection{Scalability Assessment}

The framework's performance will be tested under different scenarios, such as high vs. low renewable penetration and varying load demands, to ensure its adaptability for global contexts~\cite{b11,b22}. This will provide a comprehensive understanding of the framework's limitations and potential for future development. Interoperability testing will validate integration with existing infrastructure and third-party systems.

\section{Discussion and Future Work}

The proposed research offers a significant contribution to the field of smart energy systems by integrating predictive maintenance, affordability, and resilience into a single, cohesive framework~\cite{b5,b17}. The novelty lies in its holistic approach and the emphasis on real-world validation using hardware-in-the-loop (HIL) experiments. The project's focus on low-cost, open-source hardware and accessibility makes it particularly relevant for global application, especially in developing regions where energy access remains a critical challenge.

The results from this research are expected to provide not only a robust technological solution but also a deeper understanding of the socio-economic factors that influence the adoption of sustainable energy technologies~\cite{b12,b22}. The integration of cybersecurity considerations from the design stage provides a new model for developing resilient infrastructure that can maintain operational integrity under hostile conditions.

Future work will focus on expanding the framework to include dynamic energy trading with neighboring microgrids, integrating vehicle-to-grid (V2G) technology, and developing advanced reinforcement learning models for real-time energy management~\cite{b10,b11}. Furthermore, the research will explore the potential of using green AI principles to reduce the computational and energy footprint of the models, aligning with the broader goal of sustainability.

The development of quantum computing applications in energy system optimization could enable solution of much larger optimization problems than currently feasible. Advanced materials integration, including smart materials and nanotechnology applications, could enhance sensor capabilities while reducing costs and improving durability.

The development of a user-friendly interface will also be a key priority to ensure the technology is accessible to non-technical users in community-based microgrid projects~\cite{b17}. Policy recommendations will be developed based on research findings to support regulatory frameworks and incentive structures that can accelerate adoption of intelligent microgrid technologies.

\section{Conclusion}
\balance
This research proposal outlines a comprehensive and innovative approach to addressing the critical challenges of maintenance, affordability, and resilience in smart microgrids. By leveraging the power of AI-enhanced IoT systems and a digital twin framework, the project aims to not only optimize system performance but also to make sustainable energy more accessible and equitable~\cite{b5,b7,b17}.

The proposed methodology, which combines theoretical modeling with practical hardware validation, is designed to produce a solution that is both academically sound and commercially viable~\cite{b16}. The successful completion of this research will contribute new algorithms, an integrated framework, and a practical guide for the deployment of next-generation microgrids, ultimately accelerating the global transition toward a decentralized and sustainable energy future~\cite{b22}.

The expected outcomes include achievement of greater than 90\% fault prediction accuracy, 30--40\% reduction in operational costs, and over 95\% cybersecurity threat detection accuracy. These measurable improvements represent significant advancements while providing tangible benefits for communities adopting intelligent microgrid technologies.

\section*{Acknowledgment}
The author thanks Dr.\ Florimond Gueniat and the College of Engineering, Birmingham City University, for their valuable guidance and support. The author also acknowledges the BCU microgrid testbed technical staff for assistance with the experimental facilities.

\balance

% ====================  BIBLIOGRAPHY  ====================
\bibliographystyle{IEEEtran}   % IEEE conference style
\bibliography{references}      % <-- points to references.bib
% =========================================================

\end{document}